# Advancements in Ship Detection: Comparative Analysis of Optical and Hyperspectral Sensors

Alyazia Al Shamsi[1,], Alavikunhu Panthakkan[2], Saeed Al Mansoori[3], and Hussain Al Ahmad[4]
[1]Department of Computer Science and Engineering, American University of Sharjah(AUS), United Arab Emirates
[2,4]College of Engineering and IT, University of Dubai(UD), United Arab Emirates
[3]Mohammed Bin Rashid Space Centre (MBRSC) Dubai, United Arab Emirates
Corresponding Author: G00090034@aus.edu; apanthakkan@ud.ac.ae

*Abstract-* In marine surveillance, applications span military and civilian domains, including ship detection, marine traffic control, and disaster management. Optical and hyperspectral satellites are key for this purpose. This paper focuses on ship detection and classification techniques, particularly comparing optical and hyperspectral remote sensing approaches. It presents a comprehensive analysis of these technologies, covering feature extraction, methodologies, and their suitability for different missions. The study highlights the importance of selecting the right sensor aligned with mission objectives and conditions, aiming to improve detection accuracy through integrated strategies. The paper examines the strengths and limitations of both technologies in various maritime applications, enhancing understanding of their usability in different operational scenarios.

## I. INTRODUCTION

Maritime transportation serves as a crucial conduit for global trade and cargo movement, primarily facilitated by ships. The increase in globalization has led to a higher volume of maritime traffic, emphasizing the importance of marine security. As a response, the field of marine security has increasingly utilized the capabilities of remote sensing technology [2]. Remote sensing plays a fundamental role in surveillance, including tasks such as monitoring traffic, identifying unauthorized fishing activities, and addressing maritime pollution. By utilizing remote sensing tools, data extraction and utilization have not only improved radar systems but have also advanced the progress of electro-optical cameras and electronic support systems. This has resulted in significant progress in marine surveillance over the last decade [1]. The history of satellite-based marine surveillance dates back to the latter half of the 20th century. In the subsequent years, dedicated satellites for maritime surveillance emerged with enhanced capabilities designed to meet the challenges of monitoring vast oceanic areas. As technology continued to advance, the combination of remote sensing techniques with satellite-based communication systems and onboard processing led to the creation of more sophisticated and specialized maritime surveillance platforms. This evolution, from early Earth observation satellites to the present era of integrated and intelligent maritime surveillance, highlights the continuous progress and innovation in utilizing space technologies to protect the maritime domain.

This study delves into a comparative examination of two prominent types of sensors: optical and hyperspectral. By investigating the strengths and limitations of each technology, the research aims to offer insights to decision-makers, researchers, and practitioners in their selection of the most appropriate sensor for specific ship detection tasks. These advanced sensors play a significant role in the field of marine surveillance, enabling versatile applications in safeguarding maritime activities, preserving the environment, and supporting military operations. While optical sensors capitalize on their capacity to detect visible and near-infrared bands, hyperspectral sensors excel in capturing detailed spectral information across an extended range [3]. The synergy between these sensor types represents the interplay between established methodologies and cutting-edge innovation, providing marine surveillance with a comprehensive array of tools to address the diverse challenges presented by the expansive marine domain.

## II. LITERATURE REVIEW

This paper focuses on the exploration of ship detection using two prominent remote sensing methods commonly used in marine surveillance: optical and hyperspectral sensors. The subsequent review of relevant literature delves into significant studies in this field, discussing sensor abilities and specific data processing techniques related to these methods. The 1990s marked a significant shift with the launch of Synthetic Aperture Radar (SAR) satellites, including the European Space Agency's ERS and Envisat. These satellites excelled in operations during nighttime and through clouds, enabling continuous surveillance, oil spill identification, and ship recognition based on radar signatures [4]. Researchers like Zhu et al. [5] have made notable contributions, introducing a hierarchical ship detection model that employs a sequential approach to eliminate false alarms using shape and texture attributes. Similarly, Liu et al. [6] proposed a ship detection technique utilizing shape and contextual information. When assessing optical and hyperspectral sensors, crucial components in ship detection, comprehending their distinct uses and advantages regarding sensor capabilities and data processing becomes essential for determining their individual strengths. Both sensor types commence ship detection through a shared framework within remote sensing, encompassing three fundamental stages: image processing, target extraction, and target identification or classification [9].

*A. Optical Ship Detection Sensors*

Advancements in optical remote sensing technology have facilitated the acquisition of high-resolution imagery, enhancing spatial and optical detail available to the detection system for informed decision-making. However, manual detection of ships within high-resolution satellite images is intricate due to voluminous data processing demands and the challenge of spotting relatively small ships in vast sea areas due to image spatial resolution constraints and sparse target distribution [10]. Fig. 1 portrays the workflow of optical remote sensing image classification [10]. The processing phase is mostly coupled with Geographical Information System (GIS)-based sea-land segmentation. Which plays a significant role in expediting feature extraction by narrowing the traversal area of the core detection algorithm and enhancing detection precision by mitigating false alarms stemming from complex land texture features [11].

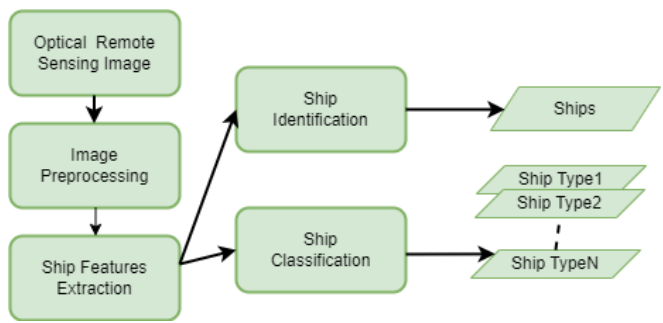

Fig. 1. A common ship detection and classification workflow

Optical ship detection sensors operate by detecting electromagnetic radiation within the visible and near-infrared segments of the electromagnetic spectrum. These sensors capture reflected sunlight from Earth's surface, allowing for the creation of images with varying levels of detail and spectral information. The key principle involves differentiating materials based on their reflectance characteristics, enabling the identification of ships amidst water. These sensors offer high spatial resolution, facilitating the detection of small objects like ships with intricate details. They primarily operate in visible and near-infrared bands, providing color information and the ability to distinguish between materials using spectral signatures. The sensor's radiometric resolution influences its capacity to discern variations in brightness levels, which in turn enhances differentiation between ships and water. However, optical sensors can be influenced by atmospheric conditions, such as haze and cloud cover, affecting detection accuracy in regions with fluctuating weather patterns.

Advantages of optical sensors encompass their well-established technology, wide sensor availability, high spatial resolution enabling detailed identification, and real-time or near-real-time data acquisition. They are well-suited for daylight operations, making them ideal for routine maritime surveillance. Nevertheless, challenges arise from phenomena like sun glint and cloud cover, which can obstruct ship features and hinder detection accuracy. Additionally, optical sensors are restricted by their reliance on sunlight, affecting their performance during nighttime or low-light conditions. While recent advancements have improved radiometric and spectral resolution, as well as atmospheric correction algorithms, challenges such as atmospheric interference, dynamic lighting conditions, and sensor calibration persist, influencing the precision of ship detection results.

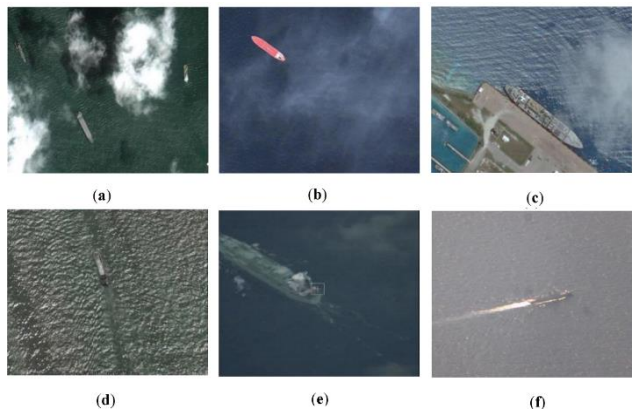

Fig.2. Some optical remote sensing ship images

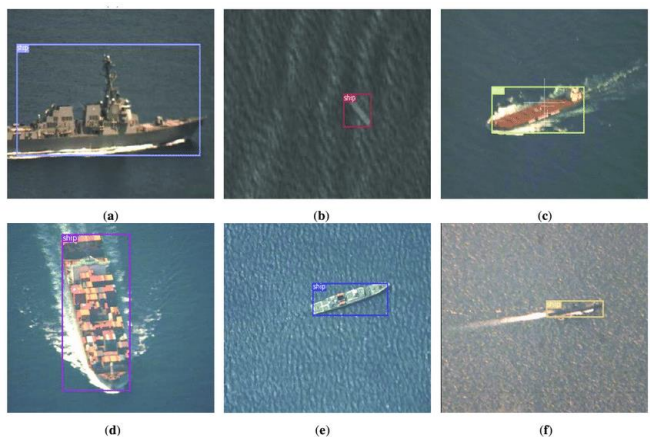

Fig.3 Ship detection results of visible light images.

### B. Hyperspectral Ship Detection Sensors

Hyperspectral sensors unravel the spectral characteristics of surfaces across numerous continuous and narrow wavelengths [12]. Hyperspectral sensor evolution traced back to NASA's 1970 study, saw continued development, including the Jet Propulsion Laboratory's (JPL) 1983 innovation and the enhanced Airborne Visible/Infrared Imaging Spectrometer (AVIRIS) in 1987 [13]. Initially, hyperspectral wasn't used within the scope of marine surveillance. It was mostly utilized in geological exploitation and vegetation studies. It's transition to marine surveillance occurred due to escalating maritime traffic that heightened ship accidents, necessitating synoptic ship monitoring under perilous circumstances, which high-resolution satellites with wide coverage provide, particularly under danger [14]. Therefore, ship detection has emerged as a pivotal facet within the remote sensing realm. Hyperspectral approaches have gained prominence in ship detection, with promising outcomes. For instance, Park et al. [15] successfully detected ships in AVIRIS images, applying a spectral matching technique to actual ship spectral library data.

Hyperspectral ship detection sensors function by capturing numerous narrow and contiguous spectral bands across the electromagnetic spectrum. This approach provides detailed spectral information for each image pixel, allowing the identification of materials, including ships, based on unique spectral signatures. These sensors offer high spectral resolution, enabling accurate material identification by distinguishing between subtle variations in reflectance. Ships possess distinctive spectral signatures due to their material composition, aiding in their differentiation from the surrounding water and even classifying specific ship types.

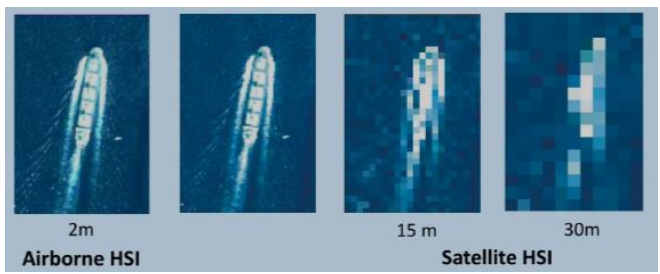

Fig.4 Ship detection results of HSI images.

Advantages of hyperspectral sensors include precise material identification, discrimination of visually similar materials with distinct spectral features, and the detection of camouflaged objects. Challenges include computational complexity in processing high-dimensional hyperspectral data, sensor cost compared to traditional optical sensors, and the need for efficient data storage and transmission methods. Recent advancements have focused on improving processing algorithms, reducing computational demands, and developing cost-effective sensors. Nonetheless, challenges related to data interpretation, rapid processing, and cost-effective implementation continue to be key areas of research and development in this field.

*C. Trade-offs between Optical and Hyperspectral Sensors*

The trade-offs between optical and hyperspectral sensors in ship detection center around spatial and spectral resolution, computational complexity, and cost-effectiveness. Optical sensors offer high spatial resolution, enabling detailed ship identification, while hyperspectral sensors provide detailed spectral information for precise material identification. The discussion addresses how hyperspectral sensors compensate for potentially lower spatial resolution with increased spectral data. The computational complexity of processing hyperspectral data can hinder real-time applications, and while optical sensors are cost-effective, hyperspectral sensors offer enhanced detection capabilities. Overall, the choice between sensor types depends on the specific maritime context and priorities.

III. SHIP DETECTION METHODS

Optical Analysis and Hyperspectral Analysis are two distinct remote sensing methods used for ship detection. Optical analysis relies on visible light and is commonly used in satellite imagery and surveillance systems for ship detection. Hyperspectral analysis involves capturing a wide range of narrow, contiguous spectral bands beyond the visible spectrum, allowing for more detailed spectral signatures of objects.

*A. Optical Analysis:*

Automated techniques are commonly integrated into optical satellites, utilizing Computer Vision methods to streamline detection procedures. Computer vision, which aims to automate tasks akin to human vision, includes object detection—a sub-domain utilizing signal processing, machine learning, and deep learning tools for identifying items within images or videos. Convolutional Neural Networks (CNNs), a type of Deep Learning model, have gained significant prominence in tasks like image recognition and object detection, as shown by diverse prior research employing various CNN variations for ship detection from satellite images [16]. A visual representation of a sample optical remote sensing image captured for detection is available in Fig. 2 [17].

Numerous advanced deep learning techniques have revolutionized ship detection in optical images, enhancing accuracy and automation. Convolutional Neural Networks (CNNs) are extensively employed due to their capacity to learn features, while Faster R-CNN efficiently locates ship regions. Real-time performance is offered by YOLO and SSD, and Mask R-CNN achieves precise instance-level segmentation. Transfer learning leverages pre-trained models, and data augmentation artificially expands datasets. Ensemble methods merge diverse architectures to enhance accuracy, and attention mechanisms improve detection in complex scenarios. Generative Adversarial Networks (GANs) generate synthetic data to mitigate scarcity. Overall, these methods have transformed ship detection by optimizing accuracy, speed, and robustness to address varied contexts and challenges.

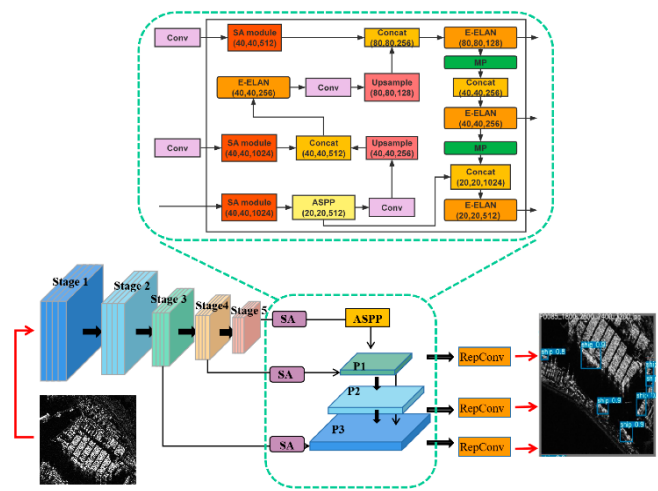

Fig.5. Flow chart of the Yolov7 based ship detection method

*B. Hyperspectral Analysis:*

Hyperspectral analysis relies on the concept of mixing models, assuming that a unit pixel comprises linear combinations of distinct signals. As sensors often capture multiple diverse signals for a single pixel, an endmember—a pure object—emerges as the pixel's composition. Various techniques, such as the N-finder algorithm (N-Findr), Pixel Purity Index (PPI), Independent Component Analysis (ICA), and VCA, are available for extracting image endmembers [14]. Previous studies in hyperspectral ship detection used arbitrary thresholds to classify ship pixels based on spectral similarity values. However, the need for adaptability to observation environments and ship types calls for an objective detection algorithm. J. Park et al.'s study serves as an illustration of this approach [14]. They employed hyperspectral remote sensing images from a hyperspectral camera to implement ship detection methods, estimate ship dimensions through data processing, and evaluate accuracy using real ship size data. The study also comprehensively applied an automated process for ship detection and determined the composite ratios contributed by endmember fractions for each pixel. This research extended to creating hyperspectral images from airborne measurements within the coastal area of the Korean Peninsula [14].

## IV. PERFORMANCE COMPARISON

In general, optical analysis is accessible and cost-effective, but it has limitations in adverse conditions. Hyperspectral analysis offers improved accuracy, especially in challenging situations, but it requires specialized equipment and may be computationally intensive. The choice between these methods depends on the specific requirements and conditions of ship detection tasks. Table 1 presents a comparative analysis of Optical and Hyperspectral methods for ship detection.

TABLE I. METHODS PEFROMANCE COMPARISON

| Criteria | Optical Analysis | Hyperspectral Analysis |
|---|---|---|
| Data Source | RGB cameras | Hyperspectral sensors |
| Weather Dependency | Prone to weather effects | Less affected by weather |
| Low-Light Operation | Ineffective at night | Effective day and night |
| Spectral Information | Limited | Extensive and detailed |
| Cost | Affordable | More expensive |
| Data Processing | Straightforward | Complex algorithms |
| Application Range | Simple recognition | Detailed ship classification |

## V. CONCLUSIONS

This paper presents a comprehensive comparative analysis of ship detection techniques, specifically focusing on optical and hyperspectral remote sensing methods. The study highlights the importance of choosing the right sensor aligned with mission objectives to enhance detection accuracy. Analyzing the strengths and limitations of these technologies across various maritime applications offers insights for decision-makers, researchers, and practitioners. The trade-offs between optical and hyperspectral sensors, considering factors like resolution, complexity, and cost, are discussed, emphasizing the need for context-specific sensor selection. The study also explores advanced ship detection methods involving Computer Vision and Deep Learning models, which have significantly improved accuracy and automation. The paper contextualizes its findings within technological progress and ethical considerations, providing practical insights, potential research directions, and policy implications. It showcases the continuous innovation in ship detection sensor technologies, contributing to the evolving field of maritime surveillance. Future scope of the project extends to various fields and applications, with its impact reaching beyond ship detection to benefit sectors like remote sensing, autonomous systems, environmental monitoring, and data-driven decision-making in both civilian and defense domains.